\documentclass[12pt]{article}

\usepackage{amsmath}

\usepackage{breqn}
\usepackage[bbgreekl]{mathbbol}
\usepackage{bbm}
\usepackage{stmaryrd} 
\usepackage[width=.8\paperwidth]{geometry}

\renewcommand{\vec}{\mathbf}
\newcommand{\rmi}{\mathrm{i}}
\newcommand{\rmd}{\mathrm{d}}
\newcommand{\tr}{\mathrm{tr}}
\newcommand{\Eref}[1]{Equation~(\ref{#1})}
\newcommand{\eref}[1]{equation~(\ref{#1})}
\newcommand{\sref}[1]{section~\ref{#1}}

\begin{document}

\title{Semi-classical Lindblad master equation\\ for spin dynamics}
\maketitle

\author{J. Dubois, Ulf Saalmann and Jan M. Rost \\
\small Max Planck Institute for the Physics of Complex Systems, N\"{o}thnitzer Stra\ss{e} 38, 01187 Dresden, Germany}
\vspace{10pt}

\begin{abstract}
We derive the semi-classical Lindblad master equation in phase space for both canonical and non-canonical Poisson brackets using the Wigner-Moyal formalism and the Moyal star-product. The semi-classical limit for canonical dynamical variables, i.\,e., canonical Poisson brackets, is the Fokker-Planck equation, as  derived before. We generalize this limit and show that it holds also for non-canonical Poisson brackets. Examples are gyro-Poisson brackets, which occur in spin ensembles, systems of recent interest in atomic physics and quantum optics. We show that the equations of motion for the collective spin variables are given by the Bloch equations of nuclear magnetization with relaxation. The Bloch and relaxation vectors are expressed in terms of the microscopic operators: The Hamiltonian and the Lindblad functions in the Wigner-Moyal formalism.
\end{abstract}

\vspace{2pc}

%
%

\section{Lindblad quantum master equation}

Since perfect isolation of a  system is an idealization~\cite{Breuer2002,Manzano2020}, open microscopic systems, which interact with an environment, are prevalent in nature as well as in experiments.
Examples include superadiance~\cite{Bhaseen2012,Munoz2019},  Cooper-pair pumping~\cite{Kamleitner2011} or  non-equilibrium nuclear processes~\cite{Antonenko1994}. In such situations  energy can be exchanged between the system and surrounding particles. The ensuing energy loss  leads to dissipation and fluctuation. A description of the dynamics which fully accounts for the interactions between the system and its environment  often implies equations which cannot be solved even numerically in a reasonable amount of time.
\par
If the relaxation time of the environment is short compared to the typical timescale of the system, a Markovian approximation can be used. As a result, the quantum dynamics of the open microscopic system described by the density matrix
$\hat{\rho} $ is governed by the Lindblad master equation~\cite{Breuer2002}
\begin{equation}
\label{eq:Lindblad_master_equation_quantum}
    \frac{\partial}{\partial t} \hat{\rho} = - \frac{\rmi}{\hbar} \left[ \hat{H} , \hat{\rho} \right] + \frac{1}{\hbar} \sum_{k=1}^{k_{\rm max}} \left( \left[ \hat{L}_k \hat{\rho} , \hat{L}^{\dagger}_k \right] + \left[ \hat{L}_k ,  \hat{\rho} \hat{L}^{\dagger}_k \right] \right) ,
\end{equation}
where  $[ \cdot , \cdot ]$ denotes the commutator. \Eref{eq:Lindblad_master_equation_quantum} has a Hamiltonian part, with the Hamilton operator $\hat{H}$, and a dissipative part that results from the interactions with the environment with Lindblad operators $\hat{L}_k$, also referred to as jump operators with arbitrarily large $k_{\rm max}$. The Lindblad master equation~\eref{eq:Lindblad_master_equation_quantum} is applied to problems in atomic physics~\cite{Buca2019,Munoz2019}, quantum optics~\cite{Manzano2016,Gardiner2000}, condensed matter~\cite{Prosen2011,Olmos2012} quantum information~\cite{Lidar1998,Kraus2008} and  decoherence~\cite{Habib1998,Brun2000}.
\par
In general, the simulation  of equation~\eref{eq:Lindblad_master_equation_quantum} in the thermodynamic limit takes extremely long time  and is numerically demanding, even with quantum-jump methods~\cite{Plenio1998}.
However, some of the quantum systems, whose dynamics is accurately described by equation~\eref{eq:Lindblad_master_equation_quantum},  exhibit classical behavior~\cite{Bhaseen2012,Munoz2019}.  Classical behavior of quantum systems often facilitates to understand  their quantum dynamics and nonlinear phenomena observed in experiments. For instance, classical trajectories allow one to identify mechanisms in microscopic systems~\cite{Corkum1993, Rost1994}, and the comparison of classical and  quantum solutions with experimental results reveals which effects observed in experiments are inherently quantum. 

A bridge to the classical phenomena can be built with semi-classical master equations derived from the quantum ones. The former can capture inherent quantum effects in the thermodynamic limit and allows us to identify connections between microscopic and macroscopic scales. For canonical variables, the semi-classical limit of equation~\eref{eq:Lindblad_master_equation_quantum} has been established \cite{Strunz1998}. Here, we formulate the Lindblad master equation~\eref{eq:Lindblad_master_equation_quantum} in the Wigner-Moyal formalism~\cite{Moyal1949} and we derive its semi-classical limit for canonical and non-canonical Poisson brackets using the Moyal-star product. 

In \sref{sec:Lindblad_quantum_master_equation} and \ref{sec:Hermitian_Lindbladian_quantum}, we recall some properties on the Lindblad quantum master equation~\eref{eq:Lindblad_master_equation_quantum}. In \sref{sec:Semiclassical_Lindblad_equation}, we formulate the Lindblad master equation~\eref{eq:Lindblad_master_equation_quantum} in the Wigner-Moyal formalism for canonical and non-canonical Poisson brackets and we derive its semi-classical limit. We show that for gyro-Poisson brackets~\cite{Ruijgrok1980}, the semi-classical Lindblad equation corresponds to the Fokker-Planck equation. This result is consistent with the result obtained for canonical Poisson brackets~\cite{Strunz1998}. In \sref{sec:Example}, as an example, we use the results obtained in \sref{sec:Semiclassical_Lindblad_equation} to derive the Bloch equations from equation~\eref{eq:Lindblad_master_equation_quantum} and  the semi-classical limit of a model for superradiance and spin squeezing. 

\subsection{Mean-field equations \label{sec:Lindblad_quantum_master_equation}}

We consider a set of quantum operators $(\hat{A}_1 , ... , \hat{A}_N)=\hat{\mathbf{A}}$. The time evolution of $\hat{\mathbf{A}}$, given by $\langle \hat{\mathbf{A}} \rangle = \tr ( \hat{\rho} \hat{\mathbf{A}} )$, is governed by
\begin{equation}
\label{eq:average_dynamics_quantum}
    \frac{\rmd}{\rmd t} \left\langle \hat{\mathbf{A}} \right\rangle = \frac{\rmi}{\hbar} \left\langle \left[ \hat{H} , \hat{\mathbf{A}} \right] \right\rangle  + \frac{1}{\hbar} \sum_{k} \left\langle \hat{L}_k^{\dagger} \left[ \hat{\mathbf{A}} , \hat{L}_k \right] - \left[ \hat{\mathbf{A}} , \hat{L}_k^{\dagger} \right] \hat{L}_k \right\rangle .
\end{equation} 
\Eref{eq:average_dynamics_quantum} can be written in the form $\rmd \langle \hat{\mathbf{A}} \rangle /\rmd t = \langle \mathbf{u} ( \hat{\mathbf{A}} ) \rangle$, where $\mathbf{u}$ is a function of the Hamilton and Lindblad operators. The latter equations does not necessarily provide a set of coupled ordinary differential equations (ODEs) for the variables $\langle \hat{A}_k \rangle$ because of correlation terms of the form $\langle \hat{A}_k ... \hat{A}_j \rangle$. In the \emph{mean-field approximation}, these terms are assumed to factorize $\langle \hat{A}_k ... \hat{A}_j\rangle=\langle \hat{A}_k \rangle ... \langle \hat{A}_j \rangle$ and, as a consequence, equation~\eref{eq:average_dynamics_quantum} becomes
\begin{equation}
\label{eq:mean_field_equations_quantum}
    \frac{\rmd}{\rmd t} \big\langle \hat{\mathbf{A}} \big\rangle = \mathbf{u} \big( \big\langle \hat{\mathbf{A}} \big\rangle \big) .
\end{equation}
Equations~\eref{eq:mean_field_equations_quantum} are referred to as the mean-field equations. They provide a set of coupled ODEs for the expectation values  $\langle \hat{A}_k \rangle$ as variables which are no longer operators. Therefore, the mean-field approximation is often referred to as the classical limit of the quantum master equations~\cite{Bhaseen2012, Munoz2019}. However, as we will show, the semi-classical limit of the quantum master equation~\eref{eq:Lindblad_master_equation_quantum} reveals  classical correlations between phase space functions of the same form as between operators in equation~\eref{eq:Lindblad_master_equation_quantum} and therefore contains more information than the mean-field equations.  Under the mean-field approximation as introduced above, quantum and semi-classical master equations reduce to the same result. 

\subsection{$\mathbb{C}$-number conjugate Lindblad operators \label{sec:Hermitian_Lindbladian_quantum}}

$\mathbb{C}$-number conjugate Lindblad (CCL) operators fulfill
\begin{equation}\label{eq:lindblad-phase}
\hat{L}_k^{\dagger} = c_k \hat{L}_k ,
\end{equation}
with $c_k \in \mathbb{C}$.  Special cases are Hermitian ($c_k = 1$) and skew-Hermitian ($c_k = -1$) Lindblad operators. In \sref{sec:Semiclassical_Lindblad_equation} we will show that for CCL operators, there is no dissipation in the classical limit. Moreover, the Lindblad quantum master equation~\eref{eq:Lindblad_master_equation_quantum} takes the form
\begin{equation}
\label{eq:Lindblad_master_equation_quantum_Hermitian}
    \frac{\partial \hat{\rho}}{\partial t} = - \frac{\rmi}{\hbar} \left[ \hat{H} ,  \hat{\rho} \right] + \frac{1}{\hbar} \sum_{k} c_k \left[ \left[ \hat{L}_k , \hat{\rho} \right] , \hat{L}_k \right] .
\end{equation}
The diffusion term in equation~\eref{eq:Lindblad_master_equation_quantum_Hermitian} is of the order $\hbar$ since $[ \hat{A} , \hat{B} ] = {\cal O}(\hbar)$. This term becomes zero  in the classical or the thermodynamic limit ($\hbar = 0$) and therefore we expect no dissipation. In addition, the semi-classical limit of the commutator $[\cdot , \cdot]$ is well known [see equation~\eref{eq:Moyal_expansion}], and therefore we can easily check from equation~\eref{eq:Lindblad_master_equation_quantum_Hermitian} that the general semi-classical Lindblad equation derived in \sref{sec:Semiclassical_Lindblad_equation} has the correct form for this specific case.

\section{Derivation of the semi-classical Lindblad equation \label{sec:Semiclassical_Lindblad_equation}}

\subsection{Reminder on the Moyal-Wigner formalism and the Moyal star-product}

The Moyal-Wigner formalism~\cite{Moyal1949} is an alternative but equivalent formulation of quantum mechanics based on a non-commutating algebra in a deformed phase space, also referred to as the deformation quantization of quantum mechanics~\cite{Blaszak2012}. In this statistical theory the operators $\hat{F}$ become scalar functions $F (\mathbf{z})$ which depend on phase-space variables $\mathbf{z} = (z_1,...,z_n)$. The state of the system is described by the quasi-probability distribution $\rho (\mathbf{z},t)$, where $\rho (\mathbf{z},t)\rmd^n\! z$ represents the probability that the system is in a small volume $\rmd^n\! z$ of phase space around $\vec z$. For a given  $\rho (\mathbf{z},t)$, the expectation value of an observable $F$ is given by
\begin{equation}
\label{eq:mean_value_observable_phase_space}
    \langle F (\mathbf{z}) \rangle = \int  \rho (\mathbf{z},t) F (\mathbf{z}) \; \rmd^n\! z  .
\end{equation}
 Also known as the \emph{Wigner quasi-probability distribution}, $\rho (\mathbf{z},t)$ is analogous to the density matrix. From the evolution of $\rho (\mathbf{z},t)$, one can determine  the time evolution of the physical observables~\eref{eq:mean_value_observable_phase_space} and also, how the quantum state evolves in phase space. 
\par
The Moyal-Wigner formalism operates in the Hilbert space of  phase-space functions with a Lie algebra. The corresponding Moyal bracket $\llbracket \cdot , \cdot \rrbracket$ is related to the standard Lie bracket $[\cdot , \cdot ]$  of quantum operators  by
\begin{equation}
\label{eq:relation_quantum_Moyal_commutator}
    \rmi \hbar \llbracket F , G \rrbracket \equiv [ \hat{F} , \hat{G} ] . 
\end{equation}
Over the Hilbert space of  phase-space functions one defines a Moyal star-product denoted with $\star$, which is associative. In terms of the Moyal star-product, the Moyal bracket between two observables in phase space $F (\mathbf{z})$ and $G (\mathbf{z})$ reads
\begin{equation}
\label{eq:Moyal_commutator_star}
    \rmi \hbar \llbracket F , G  \rrbracket = F \star G - G \star F .
\end{equation}
The Moyal bracket is also a Lie bracket~\cite{Blaszak2012},
\begin{subequations}
\begin{eqnarray}
&& \llbracket F,G \rrbracket = -\llbracket G,F \rrbracket , \label{eq:Moyal_antisymmetry} \\
&& \llbracket F,G\star H \rrbracket = \llbracket F,G \rrbracket \star H + G \star \llbracket F,H \rrbracket , \label{eq:Moyal_Leibniz} \\
&& \llbracket F,\llbracket G,H\rrbracket\rrbracket + \llbracket G,\llbracket H,F\rrbracket\rrbracket + \llbracket H,\llbracket F,G\rrbracket\rrbracket = 0 \label{eq:Moyal_Jacobi}
\end{eqnarray}
\end{subequations}
satisfying the properties of
 antisymmetry~\eref{eq:Moyal_antisymmetry}, the Leibniz's rule~\eref{eq:Moyal_Leibniz} and the Jacobi identity~\eref{eq:Moyal_Jacobi}. We note that we can also introduce the Moyal anticommutator $\rmi \hbar \llparenthesis F , G \rrparenthesis \equiv F \star G + G \star F$. Together with equation~\eref{eq:relation_quantum_Moyal_commutator} and the relation $2 \hat{F} \hat{G} = [ \hat{F} , \hat{G} ] + ( \hat{F} , \hat{G} )$, one can show that a triple product of quantum operators  reads in terms of the Moyal-Wigner functions 
\begin{equation}
\label{eq:triple_product_quantum_Moyal}
    \hat{F} \hat{G} \hat{H} \equiv F \star (G \star H) = (F \star G) \star H . 
\end{equation}
From~\eref{eq:triple_product_quantum_Moyal} it is clear that one can express a product of arbitrarily many operators in  terms of the Moyal-Wigner functions.  

\subsection{Moyal star-product for canonical dynamical variables}

The Moyal star-product has been first introduced for canonical phase-space variables~\cite{Moyal1949} $\mathbf{z} = (\mathbf{r},\mathbf{p})$ with $n = 2 m$, where $\mathbf{r} = (r_1,...,r_m)$ is, for instance, the position and $\mathbf{p} = (p_1,...,p_m)$ its canonically conjugate momentum. The Moyal star-product $\star$ for canonical dynamical variables~\cite{Moyal1949,  Blaszak2012, Littlejohn1986, Soloviev2014} reads
\begin{equation}
\label{eq:Moyal_star_product_canonical}
    \star = \exp \left[ \frac{\rmi \hbar}{2} \left( \overleftarrow{\frac{\partial}{\partial \mathbf{r}}} \cdot \overrightarrow{\frac{\partial}{\partial \mathbf{p}}} - \overleftarrow{\frac{\partial}{\partial \mathbf{p}}} \cdot \overrightarrow{\frac{\partial}{\partial \mathbf{r}}} \right) \right] ,
\end{equation}
where $\overleftarrow{\cdot}$ and $\overrightarrow{\cdot}$ refer to application of the partial derivative to the left-hand and the right-hand side of the star-product, respectively. In addition to being associative, the Moyal-star product for canonical variables given by equation~\eref{eq:Moyal_star_product_canonical} is commutative to 0th order in $\hbar$, i.e., $F \star G = F G + {\cal O}(\hbar) = G F + {\cal O}(\hbar)$, and its identity is the unity, i.e., $F\star 1 = 1 \star F = F$. Up to 2nd order in $\hbar$, the Moyal bracket corresponds to the canonical Poisson bracket, i.e., $\llbracket F,G \rrbracket = \lbrace F , G \rbrace + {\cal O}(\hbar^2)$, where
\begin{equation}
    \lbrace F,G \rbrace = \frac{\partial F}{\partial \mathbf{r}} \cdot \frac{\partial G}{\partial \mathbf{p}} - \frac{\partial F}{\partial \mathbf{p}} \cdot \frac{\partial G}{\partial \mathbf{r}} .
\end{equation}

\subsection{Moyal star-product for non-canonical dynamical variables}

As before, we consider a finite-dimensional Hamiltonian system whose (phase-space) variables are denoted as $\mathbf{z} = (z_1,...,z_n)$. It is given by a Hamiltonian $H(\mathbf{z})$ and a general Poisson bracket~\cite{Cary1983}
\begin{equation}
\label{eq:Poisson_bracket_non-canonical}
    \lbrace F , G \rbrace = \frac{\partial F}{\partial \mathbf{z}} \cdot \mathbb{J}(\mathbf{z}) \frac{\partial G}{\partial \mathbf{z}} ,
\end{equation}
where $\mathbb{J}(\mathbf{z})$ is the Poisson matrix. The Poisson bracket~\eref{eq:Poisson_bracket_non-canonical} is antisymmetric, therefore $\mathbb{J}^{\rm T}{=}{-}\mathbb{J}$, and satisfies the Jacobi identity, cf.\ also equation~\eref{eq:Moyal_Jacobi}. If the Poisson matrix is symplectic, it reads $\mathbb{J}(\mathbf{z}) = [\mathbb{0} , \mathbb{I} ; - \mathbb{I} , \mathbb{0} ]$ and the Poisson bracket is called canonical, otherwise it is non-canonical. For non-canonical dynamical variables, the Moyal star-product up to 2nd order in $\hbar$ is given by
~\cite{Kontsevich2003, Behr2004, Kupriyanov2008}
\begin{equation}
\label{eq:Moyal_star_product_non-canonical}
    \star = \exp \left[ \frac{\rmi \hbar}{2} \overleftarrow{\frac{\partial}{\partial \mathbf{z}}} \cdot \mathbb{J} (\mathbf{z}) \overrightarrow{\frac{\partial}{\partial \mathbf{z}}} \right] + \hbar^2 \tilde{\star} + {\cal O}(\hbar^3) ,
\end{equation}
where $\tilde{\star}$ is a higher order star-product, explicitly given before \cite{Kontsevich2003, Behr2004, Kupriyanov2008}. It arises because the derivatives with respect to the dynamical variables and the Poisson matrix $\mathbb{J}(\mathbf{z})$ in the exponential of equation~\eref{eq:Moyal_star_product_non-canonical} do not commute. It is symmetric, i.e., $F \tilde{\star} G = G \tilde{\star} F$ and vanishes for constant Poisson matrices, since it depends only on the derivatives of the Poisson matrix but not on the matrix itself. This is in particular true for the symplectic matrix (corresponding to the canonical Poisson bracket), for which  $\tilde{\star}$ and  higher-order corrections vanish such that~\eref{eq:Moyal_star_product_non-canonical} reduces to~\eref{eq:Moyal_star_product_canonical}.
\par
To the best of our knowledge, there is no explicit formula of the star product for non-canonical variables at all orders in $\hbar$. There is, however, a recurrence method~\cite{Kupriyanov2008} to determine the expression of the correcting terms of the non-canonical Moyal star-product at higher orders in $\hbar$ which ensures that the Moyal star-product possesses the right properties (in particular the associativity) to a given order in $\hbar$, $(F \star G) \star H = F \star (G \star H) + {\cal O}(\hbar^3)$~\cite{Kontsevich2003, Behr2004, Kupriyanov2008}: At 0th order in $\hbar$ it results from the associativity of the multiplication, at 1st order it results from the Leibniz's rule of the non-canonical Poisson bracket~\eref{eq:Poisson_bracket_non-canonical}, and at 2nd order it results from the Jacobi identity which is satisfied by the non-canonical Poisson bracket~\eref{eq:Poisson_bracket_non-canonical}. 
For our derivation of the semi-classical Lindblad equation, we use the associativity of the star-product up to the 2nd order in $\hbar$. 
\par
Since the Moyal star-product can easily be expanded in a series of $\hbar$, it is particularly well-suited for the derivation of semi-classical equations from quantum master equations. Up to 2nd order in $\hbar$, the Moyal bracket reads
\begin{equation}
\label{eq:Moyal_expansion}
    \llbracket F , G \rrbracket = \lbrace F , G \rbrace + {\cal O}(\hbar^2)\,.
\end{equation}
The 1st-order term in $\hbar$ vanishes in~\eref{eq:Moyal_expansion} due to the antisymmetry of the Poisson matrix.

\subsection{The Lindblad master equations in the Wigner-Moyal formalism and their semi-classical limit}

With equation~\eref{eq:relation_quantum_Moyal_commutator}, the commutator~\eref{eq:Moyal_commutator_star} and equation~\eref{eq:triple_product_quantum_Moyal}, we can formulate the Lindblad master equation~\eref{eq:Lindblad_master_equation_quantum} for the phase space distribution $\rho (\mathbf{z},t)$ in terms of the Moyal bracket and the Moyal star-product as
\begin{equation}
\label{eq:Lindblad_master_equation_quantum_Moyal}
    \frac{\partial \rho}{\partial t} = \llbracket H , \rho \rrbracket + \rmi \sum_{k} \bigg ( \llbracket L_k \star \rho , L^{\ast}_k \rrbracket + \llbracket L_k ,  \rho \star L^{\ast}_k \rrbracket \bigg ) ,
\end{equation}
where $H(\mathbf{z},t)$ is the Hamiltonian and $L_k (\mathbf{z},t)$ are the Lindblad functions. The scalar functions $L^{\ast}_k (\mathbf{z},t)$ are complex conjugate to $L_k (\mathbf{z},t)$. We note that the Hamiltonian and the Lindblad functions may depend explicitly on time. Equation~\eref{eq:Lindblad_master_equation_quantum_Moyal} is the Lindblad equation in the Wigner-Moyal formalism for both, canonical and non-canonical phase-space variables. Note that equation~\eref{eq:Lindblad_master_equation_quantum_Moyal} differs from a semi-classical Lindblad equation put forward by Bondar {\it et al.}~\cite{Bondar2016}. 
In our case the triple products, for instance $\hat{L}_k \hat{\rho} \hat{L}_k
^{\ast}$ in equation~\eref{eq:Lindblad_master_equation_quantum}, become $( L_k \star \rho ) \star L_k^{\ast}$ or equivalently $L_k \star ( \rho \star L_k^{\ast} )$ due to the associativity of the star product, instead of $L_k \star \rho \star L_k^{\ast}$~\cite{Bondar2016}. 
\par
Equation~\eref{eq:Lindblad_master_equation_quantum_Moyal} is in general an infinite-order partial differential equation, and can be written in a series of $\hbar$. In order to obtain the semi-classical limit of the Lindblad quantum master equation~\eref{eq:Lindblad_master_equation_quantum}, we expand the right-hand side of equation~\eref{eq:Lindblad_master_equation_quantum_Moyal} up to 1st order in $\hbar$. We obtain 
\begin{eqnarray}
\frac{\partial \rho}{\partial t} &=& \left\lbrace H ,  \rho \right\rbrace + \rmi \sum_{k} \bigg ( \left\lbrace L_k \rho , L_k^{\ast} \right\rbrace + \left\lbrace L_k , \rho L_k^{\ast} \right\rbrace \bigg ) \nonumber \\ 
    && - \dfrac{\hbar}{2} \sum_{k} \bigg ( \left\lbrace \left\lbrace L_k , \rho \right\rbrace , L_k^{\ast} \right\rbrace + \left\lbrace L_k , \left\lbrace \rho , L^{\ast}_k \right\rbrace \right\rbrace \bigg ) + {\cal O}(\hbar^2 ) .
\label{eq:semi_classical_Lindblad}
\end{eqnarray}
Equation~\eref{eq:semi_classical_Lindblad} has the same form as the equation obtained for canonical Poisson brackets~\cite{Strunz1998}, but with the important difference that the Poisson brackets have been replaced by non-canonical Poisson brackets~\eref{eq:Poisson_bracket_non-canonical}. Two main parts in equation~\eref{eq:Lindblad_master_equation_quantum_Moyal} govern the dynamics: A Hamiltonian part [1st term on the right-hand side of equation~\eref{eq:semi_classical_Lindblad}] corresponding to the Liouville equation if $L_k = 0$ for all $k$, and a part coming from the dissipative term in equation~\eref{eq:Lindblad_master_equation_quantum}. 
\par
The CCL functions corresponding to CCL operators~\eref{eq:lindblad-phase} fulfill $L_k^{\ast} = c_k L_k$. If $L_k^{\ast} = c_k L_k$ for all $k$, the 0th order in $\hbar$ on the right-hand side of equation~\eref{eq:semi_classical_Lindblad} vanishes due to the antisymmetry of the Poisson bracket. In addition, if we substitute the Moyal bracket into equation~\eref{eq:Lindblad_master_equation_quantum_Hermitian} and  use equation~\eref{eq:Moyal_expansion}, it is easy to check that the resulting expression is the same as the one obtained in equation~\eref{eq:semi_classical_Lindblad} for $L_k^{\ast} = c_k L_k$.

\subsection{Canonical and gyro-Poisson brackets: the Fokker-Planck equation}

In this section, we consider Poisson matrices such that
\begin{equation}
\label{eq:Poisson_matrix_conditions_spin_canonical}
    \frac{\partial}{\partial \mathbf{z}} \cdot \mathbb{J} (\mathbf{z}) = \mathbf{0} ,
\end{equation}
i.e., $\partial J_{ij} (\mathbf{z}) / \partial z_i = 0$ for all $i$ and $j$. This condition is fulfilled for canonical Poisson brackets, Poisson brackets which do not depend on the phase-space variables, gyro-Poisson brackets~\cite{Ruijgrok1980} as used in equation~\eref{eq:Poisson_bracket_spin_dependence} below, and Poisson brackets in Nambu systems~\cite{Nambu1973}, for instance. These brackets can be used  for  spin systems or particles driven by electromagnetic fields. Using condition~\eref{eq:Poisson_matrix_conditions_spin_canonical} and the antisymmetry of the Poisson matrix, we rewrite equation~\eref{eq:semi_classical_Lindblad} as
\begin{subequations}
\label{eq:Fokker_planck_equationALL}
\begin{equation}
\label{eq:Fokker_planck_equation}
    \frac{\partial}{\partial t} \rho (\mathbf{z},t) + \frac{\partial}{\partial z_i} \big[ u_i (\mathbf{z} , t ) \rho (\mathbf{z},t) \big] - \frac{1}{2} \frac{\partial^2}{\partial z_i \partial z_j} \big[ D_{ij} (\mathbf{z} , t ) \rho (\mathbf{z} ,t) \big] = 0 ,
\end{equation}
where the Einstein summation convention has been used. 
\Eref{eq:Fokker_planck_equation} corresponds to the \emph{Fokker-Planck equation}~\cite{Risken1984,Breuer2002}. Therefore, just as for canonical variables~\cite{Strunz1998}, the semi-classical Lindblad equation for non-canonical variables is governed by a Fokker-Planck equation, provided \eref{eq:Poisson_matrix_conditions_spin_canonical} holds. The \emph{drift vector} $\mathbf{u}(\mathbf{z},t) = ( u_1 , ... , u_n )$ and the \emph{diffusion matrix} $\mathbb{D} (\mathbf{z},t) = [D_{ij}]$ read as a function of the Hamiltonian and the Lindblad functions
\begin{eqnarray}
\label{eq:semiclassical_drift_velocity}
    u_i (\mathbf{z} , t ) &=& \left\lbrace z_i , H \right\rbrace - \rmi \sum_{k} \bigg( L_k \left\lbrace z_i , L_k^{\ast} \right\rbrace + L_k^{\ast} \left\lbrace L_k , z_i \right\rbrace \bigg) \nonumber \\
    && - \dfrac{\hbar}{2} \sum_{k} \bigg( \left\lbrace \left\lbrace L_k , z_i \right\rbrace , L_k^{\ast} \right\rbrace + \left\lbrace \left\lbrace L_k^{\ast} , z_i \right\rbrace , L_k \right\rbrace \bigg) + {\cal O}(\hbar^2)  , \\
\label{eq:semiclassical_tensor_diffusion}
    D_{ij} (\mathbf{z} , t ) &=& \hbar \sum_{k} \bigg( \left\lbrace z_i , L_k \right\rbrace \left\lbrace z_j , L_k^{\ast} \right\rbrace + \left\lbrace z_i , L_k^{\ast} \right\rbrace \left\lbrace z_j, L_k \right\rbrace \bigg) + {\cal O}(\hbar^2) . 
\end{eqnarray}
\end{subequations}
Note that the (non-Hermitian) Lindblad functions $L_k$ contribute to the drift vector also at first order in $\hbar$.
We have verified that this is also true for canonical Poisson brackets as could have been anticipated since \eref{eq:semi_classical_Lindblad} has the same form as the equation obtained for canonical Poisson brackets~\cite{Strunz1998}. This 1st-order-$\hbar$ contribution in $u_i$, however, is missing in \cite{Strunz1998}, most likely due to an omission during the quite tedious calculation.
The diffusion matrix~\eref{eq:semiclassical_tensor_diffusion} is real positive semi-definite. The Fokker-Planck equation governs the dynamics for classical stochastic Markovian systems~\cite{Breuer2002}. The 
results~(\ref{eq:Fokker_planck_equation}--\textsl{c})
we obtain from the quantum master equation are therefore consistent with  classical stochastic theory. The 2nd term in the Fokker-Planck equation~\eref{eq:Fokker_planck_equation} describes a deterministic drift. The 3rd term describes the diffusion of the stochastic variable, where $\mathbb{D}$ is known as the \emph{diffusion matrix}. The diffusion matrix $\mathbb{D} (\mathbf{z})$ is of order $\hbar$, and as a consequence the diffusion term in equation~\eref{eq:Fokker_planck_equation} corresponds to \emph{quantum fluctuations} or \emph{quantum noise}. 
The mean value of an observable $F (\mathbf{z})$ given by equation~\eref{eq:mean_value_observable_phase_space} reads
\begin{equation}
\label{eq:average_dynamics_semi_classical}
    \frac{\rmd}{\rmd t} \left\langle F (\mathbf{z}) \right\rangle = \left\langle \frac{\partial F}{\partial z_i} u_i (\mathbf{z} , t) + \frac{1}{2} \frac{\partial^2 F}{\partial z_i \partial z_j} D_{ij} \right\rangle .
\end{equation} 
In particular, the average of the dynamical variables is given by $\rmd \langle \mathbf{z} \rangle /\rmd t = \langle \mathbf{u} (\mathbf{z}) \rangle$. 
\par
In the classical limit ($\hbar = 0$), the diffusion matrix vanishes. In this limit, one gets
\begin{equation}
\label{eq:dynamical_equation_classical_limit}
    \frac{\rmd \mathbf{z}}{\rmd t} = \left\lbrace \mathbf{z} , H \right\rbrace - \rmi \sum_{k=1}^N \bigg( L_k \left\lbrace \mathbf{z} , L_k^{\ast} \right\rbrace - L_k^{\ast} \left\lbrace \mathbf{z} , L_k \right\rbrace \bigg) .
\end{equation}
Hence,  equation~\eref{eq:semi_classical_Lindblad} in the classical limit corresponds to a global description of the dynamics while equation~\eref{eq:dynamical_equation_classical_limit} corresponds to a local description of the dynamics. Both are equivalent in the sense that they carry the same amount of information. Classical dissipation is given by 
\begin{equation}
    \dfrac{\partial}{\partial \mathbf{z}} \cdot  \mathbf{u} (\mathbf{z},t) = 2 i \sum_{k=1}^{N} \lbrace L_k^{\ast} , L_k \rbrace .
\end{equation}
For CCL operators~\eref{eq:lindblad-phase} and corresponding CCL functions, the 2nd term on the right-hand side of equation~\eref{eq:dynamical_equation_classical_limit} vanishes and equation~\eref{eq:dynamical_equation_classical_limit} becomes $\mathbf{u} (\mathbf{z} , t) = \lbrace \mathbf{z} , H \rbrace$. Equation~\eref{eq:semi_classical_Lindblad} becomes Liouville's equation. The system is \emph{Hamiltonian} in the classical limit without dissipation.

\section{Examples \label{sec:Example}}

In the following, we consider two different examples of spin systems. In both cases, the dynamics of many spins is described by a collective spin variable $\mathbf{S}$~\cite{Bhaseen2012, Munoz2019}. The Poisson bracket is given by
\begin{equation}
\label{eq:Poisson_bracket_spin_dependence}
    \left\lbrace F , G \right\rbrace  = \mathbf{S} \cdot \left( \frac{\partial F}{\partial \mathbf{S}} \times \frac{\partial G}{\partial \mathbf{S}} \right) ,
\end{equation}
whose \emph{Casimir invariants} are functions of the collective spin norm $|\mathbf{S}|$.
In the first example, we consider the Lindblad quantum master equation for a model of superradiance and spin squeezing. We derive the equations for the expectation value of the spin dynamics quantum mechanically. Then, we derive the equations for the expectation value of the spin dynamics from our semi-classical equations [see equation~\eref{eq:average_dynamics_semi_classical}], and we show that  both sets of equations are of the same form. 
In the second example, we consider a general form of the Hamiltonian and the Lindblad functions. We show that the mean-field equations for the semi-classical spin dynamics are the Bloch equation with relaxation terms.

\subsection{Superradiance and spin squeezing \label{sec:supperradiance}}
Superradiance and spin squeezing can be obtained from a system of $N$ spins governed by the master equation for the reduced density matrix $\hat{\rho}$~\cite{Munoz2019}
\begin{subequations}
\begin{equation}
\label{eq:master_equation_quantum_superradiance}
\frac{\partial\hat{\rho}}{\partial t} = - \frac{i}{\hbar} \left[ \hat{H} , \hat{\rho} \right] + \dfrac{1}{\hbar} \left( \left[ \hat{L} , \hat{\rho} \hat{L}^{\dagger} \right] + \left[ \hat{L} \hat{\rho} , \hat{L}^{\dagger} \right] \right) ,
\end{equation}
where the Hamiltonian and the Lindblad operators are
\begin{eqnarray}
    &&\hat{H} = \Omega \hat{S}_x , \\
    &&\hat{L} = \sqrt{\dfrac{\Gamma}{2 J}} \left[ \hat{S}_x \left( \cos \theta + \sin \theta \right) - \rmi \hat{S}_y \left( \cos \theta - \sin \theta \right) \right] ,
\end{eqnarray}
\end{subequations}
respectively.
The parameter $\Omega$ and $\Gamma$ are the driving amplitude and the quantum-jump rate, respectively. The operator $\hat{\mathbf{S}} = ( \hat{S}_x,\hat{S}_y,\hat{S}_z )$ is the collective spin operator obeying the commutation relations $[\hat{S}_i , \hat{S}_j ] = \rmi \hbar \epsilon_{ijk} \hat{S}_k$. The driving amplitude is $\Omega$ and the total angular momentum is $J = N/2$.

\subsubsection{Mean-field equations computed quantum mechanically}
\mbox{}\\
The commutator of the spins is given by $[ \hat{S}_i , \hat{S}_j] = \rmi \hbar \epsilon_{i j k} \hat{S}_k$. From the equations of motion of the expectation values of quantum observables~\eref{eq:average_dynamics_quantum}, we obtain for the quantum Lindblad master equation~\eref{eq:master_equation_quantum_superradiance}
\begin{subequations}
\begin{eqnarray}
&& \frac{\rmd}{\rmd t} \langle \hat{S}_x \rangle = \Omega_c \langle \hat{S}_x \hat{S}_z \rangle - \dfrac{\Gamma \hbar}{2 J} \langle \hat{S}_x \rangle \left[ 1 - \sin (2 \theta) \right] , \label{eq:equations_motion_spins_sx} \\
&& \frac{\rmd}{\rmd t} \langle \hat{S}_y \rangle = - \Omega \langle \hat{S}_z \rangle + \Omega_c \langle \hat{S}_y \hat{S}_z \rangle - \dfrac{\Gamma \hbar}{2 J} \langle \hat{S}_y \rangle \left[ 1 + \sin (2 \theta) \right] , \label{eq:equations_motion_spins_sy} \\
&&\frac{\rmd}{\rmd t} \langle \hat{S}_z \rangle = \Omega \langle \hat{S}_y \rangle - \Omega_c \left( \langle \hat{S}_x^2 \rangle + \langle \hat{S}_y^2 \rangle \right) - \dfrac{\Gamma \hbar}{J} \langle \hat{S}_z \rangle , \label{eq:equations_motion_spins_sz} 
\end{eqnarray}
\end{subequations}
where $\Omega_c = \Gamma \cos (2 \theta) / J$. In the mean-field equations~\cite{Munoz2019} one assumes $\langle\hat{S}_i \hat{S}_j \rangle \approx \langle \hat{S}_i \rangle \langle \hat{S}_j \rangle$. As a result, we obtain a set of dynamical equations for the expectation value of the spin operators
\begin{subequations}
\begin{eqnarray}
&& \frac{\rmd}{\rmd t} \langle \hat{S}_x \rangle = \Omega_c \langle \hat{S}_x \rangle \langle \hat{S}_z \rangle - \dfrac{\Gamma \hbar}{2 J} \langle \hat{S}_x \rangle \left[ 1 - \sin (2 \theta) \right] , \label{eq:equations_motion_spins_mean_field_sx} \\
&& \frac{\rmd}{\rmd t} \langle \hat{S}_y \rangle = - \Omega \langle \hat{S}_z \rangle + \Omega_c \langle \hat{S}_y \rangle \langle \hat{S}_z \rangle - \dfrac{\Gamma \hbar}{2 J} \langle \hat{S}_y \rangle \left[ 1 + \sin (2 \theta) \right] , \label{eq:equations_motion_spins_mean_field_sy} \\
&& \frac{\rmd}{\rmd t} \langle \hat{S}_z \rangle = \Omega \langle \hat{S}_y \rangle - \Omega_c \left( \langle \hat{S}_x \rangle^2 + \langle \hat{S}_y \rangle^2 \right) - \dfrac{\Gamma \hbar}{J} \langle \hat{S}_z \rangle .  \label{eq:equations_motion_spins_mean_field_sz}
\end{eqnarray}
\end{subequations}

\subsubsection{Expectation values from the semi-classical Lindblad equation}
\mbox{}\\
In the semi-classical approach, the operators become scalar functions of dynamical phase-space variables. The Hamiltonian operator $\hat{H}$ becomes $H (\mathbf{S})$, and the Lindblad operators $\hat{L}$ become $L (\mathbf{S})$. In this case, they read
\begin{subequations}
\begin{eqnarray}
    && H(\mathbf{S}) = \Omega S_x , \label{eq:Hamiltonian_MoyalFunctions_superradiance} \\
    && L (\mathbf{S}) = \sqrt{\dfrac{\Gamma}{2 J}} \left[ S_x \left( \cos \theta + \sin \theta \right) - \rmi S_y \left( \cos \theta - \sin \theta \right) \right] . \label{eq:Lindblad_MoyalFunctions_superradiance}
\end{eqnarray}
\end{subequations}
In equation~\eref{eq:average_dynamics_semi_classical}, we substitute the gyro-Poisson bracket~\eref{eq:Poisson_bracket_spin_dependence} and we use $F = \mathbf{S}$. We obtain
\begin{subequations}
\begin{eqnarray}
&&\frac{\rmd}{\rmd t} \langle S_x \rangle  = \Omega_c \langle S_x S_z \rangle - \dfrac{\Gamma \hbar}{2 J} \langle S_x \rangle \left[ 1 - \sin (2 \theta) \right] , \label{eq:equations_motion_spins_sc_sx} \\
&&\frac{\rmd}{\rmd t} \langle S_y \rangle = - \Omega \langle S_z \rangle + \Omega_c \langle S_y S_z \rangle - \dfrac{\Gamma \hbar}{2 J} \langle S_y \rangle \left[ 1 + \sin (2 \theta) \right] , \label{eq:equations_motion_spins_sc_sy} \\
&&\frac{\rmd}{\rmd t} \langle S_z \rangle = \Omega \langle S_y \rangle - \Omega_c \left( \langle S_x^2 \rangle + \langle S_y^2 \rangle \right) - \dfrac{\Gamma \hbar}{J} \langle S_z \rangle ,  \label{eq:equations_motion_spins_sc_sz}
\end{eqnarray}
\end{subequations}
which have the same form as
equations~(\ref{eq:equations_motion_spins_sx}--\textsl{c})
derived from the quantum-mechanical approach. Beyond the mean-field equations, our semi-classical approach takes into account the semi-classical spin correlations $\langle \hat{S}_i \hat{S}_j \rangle$. Yet  the long-time evolution of the expectation values according to 
equations~(\ref{eq:equations_motion_spins_sx}--\textsl{c})
and 
equations~(\ref{eq:equations_motion_spins_sc_sx}--\textsl{c})
differ due to these correlation terms which involve operators and functions, respectively. However, in  mean-field approximation, these differences in the correlations is suppressed and quantum as well as semi-classical dynamics is governed by the same mean-field equations. 

\subsection{Collective spin systems and the Bloch equations of nuclear magnetization \label{sec:Bloch_equations}}
We consider a general form of the Hamiltonian and Lindblad functions, and a collective spin variable $\mathbf{S}$ with a Poisson bracket~\eref{eq:Poisson_bracket_spin_dependence}. The expectation value of the collective spin variables is given by equation~\eref{eq:average_dynamics_semi_classical}. In this case, $\rmd \langle \mathbf{S} \rangle / \rmd t = \langle \mathbf{u}(\mathbf{S}) \rangle$, it reads
\begin{eqnarray*}
    \frac{\rmd}{\rmd t} \langle  \mathbf{S} \rangle &=& \bigg\langle  \mathbf{S} \times \left[ - \frac{\partial H}{\partial \mathbf{S}} + \rmi \sum_{k} \left( L_k \frac{\partial L^{\ast}_k}{\partial \mathbf{S}} - L^{\ast}_k \frac{\partial L_k}{\partial \mathbf{S}} \right) \right] \nonumber \\
	&& + \dfrac{\hbar}{2} \sum_{k} \left( \left\lbrace L_k , \mathbf{S} \times \frac{\partial L^{\ast}_k}{\partial \mathbf{S}} \right\rbrace + \left\lbrace L_k^{\ast} , \mathbf{S} \times \frac{\partial L_k}{\partial \mathbf{S}} \right\rbrace \right) \bigg\rangle + {\cal O}(\hbar^2) .
\end{eqnarray*} 
In case of a  mean-field scenario, where the correlation function almost vanishes, i.e., for any function $f$, $\langle f( \mathbf{S} ) \rangle \approx f (\langle \mathbf{S} \rangle )$, the expectation value of the collective spin variable  is given by a Bloch equation with relaxation terms~\cite{Bloch1946}
\begin{subequations}
\begin{equation}
\label{eq:Bloch_equations}
    \frac{\rmd}{\rmd t} \langle  \mathbf{S} \rangle = \langle  \mathbf{S} \rangle \times \mathbf{B} (\langle  \mathbf{S} \rangle) - \mathbf{R} (\langle  \mathbf{S} \rangle) ,
\end{equation}
where $\mathbf{B}$ is the Bloch vector and $\mathbf{R}$ is the relaxation term. They are related to the Hamiltonian and the Lindblad functions through
\begin{eqnarray}
\label{eq:Bloch_Bloch_vector_semiclassical_general_expression}
   && \mathbf{B} (\mathbf{S}) = - \frac{\partial H}{\partial \mathbf{S}} + \rmi \sum_{k} \left( L_k \frac{\partial L^{\ast}_k}{\partial \mathbf{S}} - L^{\ast}_k \frac{\partial L_k}{\partial \mathbf{S}} \right)  - \frac{\hbar}{2} \sum_{k} \left[ \mathbf{v} (L_k , L_k^{\ast} ) + \mathbf{v}  (L_k^{\ast} , L_k ) \right] , \\
\label{eq:Bloch_relaxation_vector_semiclassical_general_expression}
   && \mathbf{R} (\mathbf{S}) = \dfrac{\hbar}{2} \sum_{k} \left[ 2 \mathbf{S} \left( \dfrac{\partial L_k}{\partial \mathbf{S}} \cdot \dfrac{\partial L_k^{\ast}}{\partial \mathbf{S}} \right) - \dfrac{\partial L_k}{\partial \mathbf{S}} \left( \mathbf{S} \cdot \dfrac{\partial L_k^{\ast}}{\partial \mathbf{S}} \right) - \dfrac{\partial L_k^{\ast}}{\partial \mathbf{S}} \left( \mathbf{S} \cdot \dfrac{\partial L_k}{\partial \mathbf{S}} \right) \right] ,
\end{eqnarray}
where
\begin{equation}
    v_i ( F , G ) = \mathbf{S} \cdot \left( \dfrac{\partial^2 F}{\partial \mathbf{S} \partial S_i} \times \dfrac{\partial G}{\partial \mathbf{S}} \right) .
\end{equation}
\end{subequations}
Both, $\mathbf{B}$ and $\mathbf{R}$, can depend on the collective spin variable and time through the Hamiltonian and Lindblad functions. The relaxation term $\mathbf{R}$ is of order $\hbar$, and therefore  relaxation is here a quantum feature. In the classical limit ($\hbar = 0$), there is no relaxation, and therefore the norm of the spin variable is conserved, i.e., $\rmd |\mathbf{S}| / \rmd t = 0$. However,  dissipation is still possible:
If the Lindblad functions are not CCL functions, there can be attractors in the dynamical system. The relaxation term also vanishes if $\partial L_k / \partial \mathbf{S} \propto \mathbf{S}$, implying that each $L_k$ is a function of the norm of the spin $|\mathbf{S}|$. In this case, $L_k$ are Casimir invariants and as a consequence the system is Hamiltonian [no dissipation term in equation~\eref{eq:semi_classical_Lindblad}]. 
\par
 The model in the first example (see \sref{sec:supperradiance}) can serve as  an application of equations~\eref{eq:Bloch_equations}.  We substitute the Hamiltonian~\eref{eq:Hamiltonian_MoyalFunctions_superradiance} and Lindblad functions~\eref{eq:Lindblad_MoyalFunctions_superradiance} into equations~\eref{eq:Bloch_Bloch_vector_semiclassical_general_expression} and~\eref{eq:Bloch_relaxation_vector_semiclassical_general_expression} to obtain
\begin{subequations}
\begin{eqnarray*}
    &&\mathbf{B} (\mathbf{S}) = - \Omega \mathbf{e}_x + \Omega_c \mathbf{S} \times \mathbf{e}_z , \\
    &&\mathbf{R} (\mathbf{S}) = \dfrac{\Gamma \hbar}{2 J} \left[ S_x (1-\sin 2\theta)  \mathbf{e}_x + S_y (1+\sin 2\theta)  \mathbf{e}_y + 2 S_z \mathbf{e}_z \right] .
\end{eqnarray*}
\end{subequations}
Using these expressions in equation~\eref{eq:Bloch_equations}, we obtain the same equations of motion for $\langle \mathbf{S} \rangle$ as 
equations~(\ref{eq:equations_motion_spins_mean_field_sx}--\textsl{c}).

\section{Conclusions}
To conclude, we have formulated the Lindblad quantum master equation~\eref{eq:Lindblad_master_equation_quantum_Moyal} in the Wigner-Moyal formalism and we have derived its semi-classical limit for canonical and non-canonical Poisson Poisson brackets. We have shown that for Poisson matrices which fulfill condition~\eref{eq:Poisson_matrix_conditions_spin_canonical}, the semi-classical limit of the Lindblad quantum master equation is a Fokker-Planck equation [see equations~\eref{eq:Fokker_planck_equation}~\eref{eq:semiclassical_drift_velocity}~\eref{eq:semiclassical_tensor_diffusion}]. Since condition~\eref{eq:Poisson_matrix_conditions_spin_canonical} includes as a special case the canonical Poisson bracket our results agree with those obtained by Strunz {\it et al.}~\cite{Strunz1998} in the canonical case.  Condition~\eref{eq:Poisson_matrix_conditions_spin_canonical} is also satisfied by gyro-Poisson brackets, such as the Poisson bracket~\eref{eq:Poisson_bracket_spin_dependence}, which occurs in spin ensembles~\cite{Buca2019,Munoz2019} and for particles in an electromagnetic field~\cite{Cary1983}.

More specifically, we have shown that the semi-classical limit of a spin ensemble whose dynamics is driven by a Lindblad master equation~\eref{eq:Lindblad_master_equation_quantum} is related to the Bloch equations with relaxations (see \sref{sec:Bloch_equations}), and we have expressed the Bloch vector and the relaxation vector of the Bloch equations as a function of the Hamiltonian and Lindblad functions in the Wigner-Moyal
formalism~(\ref{eq:Bloch_Bloch_vector_semiclassical_general_expression},\,\textsl{c}).

To illustrate the relation between the mean-field approximation and our semi-classical approach, we have applied the semi-classical limit of the Lindblad quantum master equation to spin ensembles  with a gyro-Poisson bracket~\eref{eq:Poisson_bracket_spin_dependence}.  While the mean-field equations obtained from the quantum-mechanical 
approach~(\ref{eq:equations_motion_spins_mean_field_sx}--\textsl{c})
agree with those obtained from the semi-classical approach, our semi-classical approach without further approximation [see
equations~(\ref{eq:Fokker_planck_equation}--\textsl{c})] 
also provides information on the spin correlation and  dissipation, even in its classical limit ($\hbar = 0$). This suggests that our semi-classical approach provides more information in the classical limit than the  traditional classical limit of the quantum formulation~\eref{eq:mean_field_equations_quantum} via
mean-field approximation \cite{Bhaseen2012, Moodie2018, Munoz2019}.

\section*{Acknowledgments}
JD thanks Christian Johansen and Alexander Eisfeld for helpful discussions.

\bibliographystyle{iopart-num}

\providecommand{\newblock}{}

\end{document}